\documentclass[aps,prx,reprint,superscriptaddress,longbibliography]{revtex4-1}

\usepackage{amsmath,amsfonts,bm,graphicx,verbatim,mathrsfs,units}
\usepackage{color,colordvi}
\definecolor{blue}{rgb}{0,0,1}
\definecolor{grey}{rgb}{0.6,0.6,0.6}

\definecolor{myurlcolor}{rgb}{0,0,0.7}
\definecolor{myrefcolor}{rgb}{0.8,0,0}
\usepackage[unicode=true,pdfusetitle, bookmarks=false,bookmarksnumbered=false,
bookmarksopen=false, breaklinks=false,pdfborder={0 0 0},backref=false,
colorlinks=true, linkcolor=myrefcolor,citecolor=myurlcolor,urlcolor=myurlcolor]{hyperref}

\newcommand{\ket}[1]{\left| #1 \right \rangle}
\newcommand{\bra}[1]{\left \langle #1 \right|}

\begin{document}

%Title and abstract----------

\title{Optimal work extraction from quantum states by photo-assisted Cooper pair tunneling}

\author{Niels L\"orch}
\affiliation{Department of Physics, University of Basel,
  Klingelbergstrasse 82, CH-4056 Basel, Switzerland}
\author{Christoph Bruder}
\affiliation{Department of Physics, University of Basel,
  Klingelbergstrasse 82, CH-4056 Basel, Switzerland}
\author{Nicolas Brunner}
\affiliation{Department of Applied Physics, University of Geneva, Chemin de Pinchat 22, 1211 Geneva, Switzerland.}
\author{Patrick P. Hofer}
\email{patrick.hofer@teorfys.lu.se}
\affiliation{Department of Applied Physics, University of Geneva, Chemin de Pinchat 22, 1211 Geneva, Switzerland.}
\affiliation{Physics Department and NanoLund, Lund University, Box 118,  22100 Lund, Sweden.}

%\thanks{}

\date{\today}

\begin{abstract}
The theory of quantum thermodynamics predicts fundamental bounds on work extraction from quantum states. As these bounds are derived in a very general and abstract setting, it is unclear how relevant they are in an experimental context, where control is typically limited. Here we address this question by showing that optimal work extraction is possible for a realistic engine. The latter consists of a superconducting circuit, where a $LC$-resonator is coupled to a Josephson junction. The oscillator state fuels the engine, providing energy absorbed by Cooper pairs, thus producing work in the form of an electrical current against an external voltage bias. We show that this machine can extract the maximal amount of work from all Gaussian and Fock states. Furthermore, we consider work extraction from a continuously stabilized oscillator state. In both scenarios, coherence between energy eigenstates is beneficial, increasing the power output of the machine. This is possible because the phase difference across the Josephson junction provides a phase reference.
\end{abstract}

%\keywords{}

\maketitle

%Main text--------------

\section{Introduction}

The desire to convert energy from the disordered form of heat into the useful form of work was one of the principal driving forces in developing the theory of thermodynamics \cite{weinberger:2013}. Recent experimental advancements in control over nano-scale systems motivated the investigation of similar processes in the regime where quantum effects start to play a role, an effort that led to the continuing development of the theory of quantum thermodynamics \cite{kosloff:2013,pekola:2015,vinjanampathy:2016,goold:2016,anders:2017}. A part of this theory which is of particular practical importance is the study of quantum thermal machines \cite{kosloff:2014,gelbwaser:2015,goold:2016,benenti:2017}. These are devices which make use of thermal gradients to perform useful tasks such as the production of work \cite{kosloff:1984,quan:2007,brunner:2012,abah:2012,sothmann:2015,hofer:2015,hofer:2016prb,roulet:2017,hardal:2017,benenti:2017,josefsson:2017,roulet:2018}, the refrigeration of a quantum degree of freedom \cite{palao:2001,linden:2010prl,levy:2012,hofer:2016,mitchison:2016,maslennikov:2017}, the creation of entanglement \cite{brask:2015njp,boyanovsky:2017,tavakoli:2018}, the determination of low temperatures \cite{hofer:2017}, or the design of thermal transistors \cite{joulain:2016} and autonomous quantum clocks \cite{erker:2017,woods2016}. 

The most prominent examples of macroscopic thermal machines are heat
engines which convert heat into work for, e.g., powering cars and airplanes.
While the distinction between heat and work seems to
be clear in these macroscopic engines where heat is produced by
combustion and the work produces a directional motion of a vessel, the
situation is much more blurry in the quantum regime. Therefore there
is an ongoing debate on the definition of work in quantum
thermodynamics
\cite{talkner:2007,horodecki:2013,frenzel:2014,skrzypczyk:2014,binder:2015,talkner:2016,gallego:2016,perarnau:2017,seah:2018}. In
spite of these arguments, it seems clear that any definition of work
needs to entail a certain level of \textit{usefulness} and
\textit{accessibility} or \textit{measurability}. Indeed, already in
classical thermodynamics, heat is defined as the energy change in the
microscopic, inaccessible degrees of freedom while work is defined as
the energy change in the degrees of freedom that are macroscopic and accessible. Such macroscopic, accessible degrees of freedom can be measured in a macroscopic measurement and do not average to zero over macroscopic time- and length-scales \cite{callen:book}. In quantum systems, all degrees of freedom are in principle microscopic and their accessibility or measurability is largely dependent on the experimental setup. Defining work as the energy change in the accessible degrees of freedom thus necessarily leads to a definition which differs from experiment to experiment.

\begin{figure}
\centering
\includegraphics[width=\columnwidth]{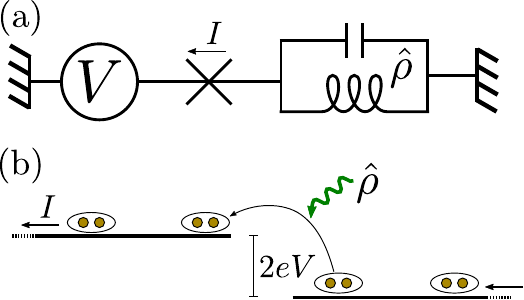}
\caption{${\rm (a)}$ Sketch of the engine. A superconducting circuit consisting of an $LC$-resonator that is coupled to a Josephson junction. {The fuel of the engine is provided by the quantum state $\hat{\rho}$, localized in the resonator.} Energy stored in the quantum state can drive an electrical current ($I$) against an external voltage bias ($V$), creating work. ${\rm (b)}$ Illustration of the work extraction mechanism. Work is created by moving Cooper pairs across a voltage-biased Josephson junction. The work is given by the energy increase of the Cooper pairs and is extracted from the quantum state $\hat{\rho}$.}
  \label{fig:schematics}
\end{figure}

{Even though the above discussion implies that work should be defined with an experimental setup in mind, fundamental limits can be obtained. The question of how much work can be extracted from a quantum state has received a lot of attention recently \cite{allahverdyan:2004,alicki:2013,perarnau:2015,aberg:2014,malabarba:2015,lostaglio:2015,korzekwa:2016,brandao:2013,horodecki:2013,skrzypczyk:2014,uzdin:2015,richens:2016,kammerlander:2016,kwon:2018,sparaciari:2017pra,elouard:2017,seah:2018} and lies at the heart of the resource theory of quantum thermodynamics  \cite{janzing:2000,horodecki:2003,brandao:2013,horodecki:2013,skrzypczyk:2014,brandao:2015,gour:2015,cwiklinski:2015,lostaglio:2015,lostaglio:2015prx,goold:2016,sparaciari:2017pra,yunger:2017}. As these works consider very abstract and general settings (hence assuming essentially full control over the system) it is not clear that the fundamental bounds are relevant in an experimental setting. In fact, recent works investigated work extraction with restricted control \cite{perry:2015}, and showed that the performance can be strongly affected by the level of control \cite{brown:2016,clivaz:2017,friis:2018,lostaglio:2018}.
It remains thus unclear whether fundamental bounds can be reached in an experimental setting.

In the present work, we address this question and show that optimal work extraction is possible in a realistic model of an engine which extracts energy from quantum states {(not to be confused with a \textit{heat} engine, as there is no heat flow that fuels the engine). Here, optimal work extraction means that the theoretical maximum given in Eq.~\eqref{eq:wmaxun} below can be obtained.}
The engine is based on a superconducting circuit and it is fueled by a quantum state localized in an $LC$-resonator (see Fig.~\ref{fig:schematics}). Work is defined in the usual way for thermoelectric devices \cite{sothmann:2015,hofer:2015,hofer:2016prb,benenti:2017,josefsson:2017}: The power is defined through the measurable electrical current that flows against an external voltage. Work is then given by the time-integral of the power which is determined by the charge transferred through the system. An electrical current against an external voltage is useful as it can in principle be used to charge a battery.} We find that {in this realistic scenario, fundamental bounds on work extraction can be reached for a number of quantum states (all Gaussian states and all Fock states). These results are also extended to the multi-mode case.}
%XXX can it really all be extracted, or are there restrictions due to the approximations we make. should we qualify this sentence? PH: Of course in an actual experiment, there are always imperfections. However, in the considered limits, all work can be extracted. Limitations should be mentioned further down.
 We note that the superconducting phase difference across a Josephson junction acts as a phase reference which is necessary to extract the energy that is stored in the coherence between energy eigenstates \cite{aberg:2014,malabarba:2015,lostaglio:2015,korzekwa:2016,kwon:2018}. {We find that such coherences can increase the power output of the engine.}
 
 In addition to considering single quantum states, we investigate work extraction from a state that is continuously being stabilized. {In this case, the engine becomes autonomous as no time-dependent control is necessary. Such a mode of operation can be implemented experimentally with current technology \cite{hofheinz:2011,westig:2017,jebari:2018}, opening up the possibility of near-future experiments on work extraction.} As a concrete example, we consider the situation where the engine converts the power provided by a laser into electrical power. Although in this example work is converted into work, we find a trade-off between the amount of extracted power and the efficiency of the process in analogy to heat engines. 

The rest of this paper is organized as follows: In Sec.~\ref{sec:machine}, we introduce the engine, its working principle and the approximations made in modeling the proposed experimental setup. Section \ref{sec:limits} reviews fundamental limits in work extraction which will be used as a benchmark. The performance of our engine is discussed in detail in Sec.~\ref{sec:results}. The paper concludes with an outlook in Sec.~\ref{sec:conclusions}.

\section{The Engine}
\label{sec:machine}
The system we consider as an engine for extracting work is depicted in Fig.~\ref{fig:schematics} and consists of a superconducting $LC$-resonator coupled to a dc-biased Josephson junction. This system is described by the Hamiltonian \cite{saidi:2001} ($\hbar=1$ throughout the paper)
\begin{equation}
\label{eq:hamtot}
\hat{H}'=\Omega\hat{a}^\dagger\hat{a}+2eV\hat{N}+E_C\hat{N}^2-E_J\cos\left(\hat{\theta}+\hat{\varphi}\right).
\end{equation}
The first term in the Hamiltonian denotes the energy of the microwave resonator, where $\hat{a}$ is the photon annihilation operator. The second and third terms denote the energy of the Cooper pairs, where $\hat{N}$ is the number of excess Cooper pairs on the side of the Josephson junction with higher chemical potential. The term linear in $\hat{N}$ corresponds to the increase in energy when Cooper pairs tunnel against the external voltage $V$. Throughout the paper, $e$ denotes the electronic charge and ${\rm e}$ denotes Euler's number. The quadratic term corresponds to the repulsive interaction of Cooper pairs which is determined by the charging energy $E_C=2e^2/C$, where $C$ denotes the capacitance of the junction. We note that any charge offset can be absorbed into the external bias voltage. %XXX please explain. PH: charge offset adds a linear term by changing the charging term to E_C(\hat{N}-N_0)^2. The linear term can be absorbed in the voltage.
 The last term in the Hamiltonian, the Josephson term, couples the tunneling Cooper pairs to the photons in the resonator. The superconducting phase operator $\hat{\theta}$ can be expressed in the charge eigenbasis as \cite{ingold:1992,gramich:2013}
\begin{equation}
\label{eq:phastun}
{\rm e}^{i\hat{\theta}}=\sum_{N=-\infty}^{\infty}|N+1\rangle\langle N|,
\end{equation}
illustrating its role in moving Cooper pairs across the junction. {The voltage fluctuations of the microwave resonators contribute to the phase across the junction through the operators} \cite{armour:2013,gramich:2013}
\begin{equation}
\label{eq:phiho}
\hat{\varphi}=2\lambda(\hat{a}+\hat{a}^\dagger),
\end{equation}
where $\lambda=\sqrt{\pi e^2Z/h}$ (where we reinstated the Planck constant), here $Z$ denotes the impedance of the microwave resonator. We note that if $\hat{\varphi}$ can be replaced by a classical variable $\varphi(t)$, a unitary transformation of the Hamiltonian (with $\hat{U}=\exp[{-i\varphi(t)\hat{N}}]$) removes $\varphi(t)$ from the Josephson term and results in the usual capacitive coupling $2e\hat{N}(V+V_{\Omega})$, where $2eV_{\Omega}=\partial_t\varphi(t)$ \cite{ingold:1992,gramich:2013}.

The goal of the engine is to extract energy stored in the resonator to drive charge (in the form of Cooper pairs) against the external voltage bias. We thus call the resonator the \textit{system} which {provides the fuel of the engine and} is governed by the Hamiltonian 
\begin{equation}
\label{eq:hsys}
\hat{H}_S=\Omega\hat{a}^\dagger\hat{a}.
\end{equation}
The work that is extracted from the system is transferred to the Cooper pairs which we denote as the work storage device governed by the second and the third terms in $\hat{H}'$. Here we are interested in the situation where the superconductors are large, such that the charging energy becomes vanishingly small. The work storage device Hamiltonian can then be approximated by
\begin{equation}
\label{eq:hwsd}
\hat{H}_w=2eV\hat{N}.
\end{equation}
This corresponds to an infinite ladder. Such a Hamiltonian has been used in multiple theoretical studies to describe a work storage device \cite{brunner:2012,aberg:2014,skrzypczyk:2014,malabarba:2015,richens:2016}.
 Of course such an unbounded Hamiltonian is in principle unphysical. However, in experimental situations, the circuit is connected to external wiring such that the excess charges can leave the system, preventing a large buildup of charge and keeping the chemical potential in the superconductor fixed. For large superconductors, the system is then well described by Eq.~\eqref{eq:hamtot} with $E_C=0$ (formally, this corresponds to superconductors of infinite size) \cite{tinkham:book}. 
When using the system to charge a battery, the excess charges are collected in a finite amount of space and charging effects have to be taken into account eventually. As we consider work extraction in a proof-of-principle manner, we focus on the regime where we can neglect $E_C$. 

We can now define a power operator which gives the change in energy in the work storage device
\begin{equation}
\label{eq:power}
\hat{P}'=-i[\hat{H}_w,\hat{H}']=\hat{I}'V=-2eVE_J\sin\left(\hat \theta+\hat{\varphi}\right),
\end{equation}
where the current operator is defined in the usual manner as $\hat{I}'=-2ei[\hat{N},\hat{H}']$. Here positive power denotes a current against the external voltage bias. We note that only if we neglect $E_C$ does the power operator reduce to the usual definition for electronic systems with a fixed voltage.

Finally, the interaction between the system and the work storage device is given by the Josephson Hamiltonian
\begin{equation}
\label{eq:hamint}
\hat{H}'_{int}=-E_J\cos\left(\hat{\theta}+\hat{\varphi}\right).
\end{equation}
This is a complicated interaction which does not conserve energy. However, since we are only interested in low-frequency (as compared to $\Omega$) observables, we can perform a rotating-wave approximation to obtain an energy-conserving interaction. To this end, we transform the Hamiltonian into a rotating frame using
\begin{equation}
\label{eq:urot}
\hat{U}_r={\rm e}^{i\hat{a}^\dagger\hat{a}\Omega t}{\rm e}^{i2e\hat{N}Vt},
\end{equation}
resulting in the Hamiltonian
\begin{equation}
\label{eq:hrot}
\begin{aligned}
\hat{H}_r=&-\frac{E_J}{2}{\rm e}^{i\hat{\theta}}{\rm e}^{i2eVt}\exp\left[2i\lambda\left(\hat{a}^\dag {\rm e}^{i\Omega t}+\hat{a} {\rm e}^{-i\Omega t}\right)\right]\\&+H.c.
\end{aligned}
\end{equation}
We now expand the displacement operator as
\begin{equation}
\label{eq:expdispl}
\begin{aligned}
&\exp\left[2i\lambda\left(\hat{a}^\dag {\rm e}^{i\Omega t}+\hat{a} {\rm e}^{-i\Omega t}\right)\right]\\&=\sum\limits_{k=0}^{\infty}i^k(\hat{a}^\dag)^k\hat{A}(k){\rm e}^{ik\Omega t}+\sum\limits_{k=1}^{\infty}i^k\hat{A}(k)\hat{a}^k{\rm e}^{-ik\Omega t},
\end{aligned}
\end{equation}
where we introduced the Hermitian operators
\begin{equation}
\label{eq:aops}
\hat{A}(k)=(2\lambda)^k{\rm e}^{-2\lambda^2}\sum\limits_{n=0}^{\infty}\frac{n!}{(n+k)!}L_{n}^{(k)}(4\lambda^2)|n\rangle\langle n |,
\end{equation}
with the generalized Laguerre polynomials $L_n^{(k)}(x)$. Setting the external voltage to
\begin{equation}
\label{eq:voltagek}
2eV=k\Omega,
\end{equation}
and dropping all the terms in the Hamiltonian which oscillate as a function of time, we obtain
\begin{equation}
\label{eq:hint}
\hat{H}_{int}=-\frac{E_J}{2}\bigg[i^k\hat{A}(k)\hat{a}^k{\rm e}^{i\hat{\theta}}+(-i)^k(\hat{a}^\dag)^k\hat{A}(k){\rm e}^{-i\hat{\theta}}\bigg].
\end{equation}
This is a resonant interaction as can be seen from
\begin{equation}
\label{eq:resonant}
[\hat{H}_{int},\hat{H}_S+\hat{H}_w]=0.
\end{equation} 
We note that in the rotating frame, $\hat{H}_{int}$ is the total Hamiltonian, while in the laboratory frame we have
\begin{equation}
\label{eq:hlab}
\hat{H}_l=\hat{H}_S+\hat{H}_w+\hat{H}_{int},
\end{equation}
where the subscript $l$ denotes the laboratory frame and the terms on the right-hand side are given in Eqs.~\eqref{eq:hsys}, \eqref{eq:hwsd}, and \eqref{eq:hint}.

Since we are only interested in the dc current, we can use a similar approximation on the current operator [cf.~Eq.~\eqref{eq:power}] which results in
\begin{equation}
\label{eq:curr}
\hat{I}=ieE_J\left[i^k\hat{A}(k)\hat{a}^k{\rm e}^{i\hat{\theta}}-(-i)^k(\hat{a}^\dag)^k\hat{A}(k){\rm e}^{-i\hat{\theta}}\right].
\end{equation}
In both Eqs.~\eqref{eq:hamint} and \eqref{eq:curr}, the first term corresponds to a Cooper pair tunneling against the voltage bias, absorbing $k$ photons from the resonator, and the second term corresponds to a Cooper pair tunneling with the voltage bias, emitting $k$ photons. We note that the definition of the current in Eq.~\eqref{eq:curr} is consistent with the definitions $\hat{I}=-2ei[\hat{N},\hat{H}_l]$ and $\hat{P}=\hat{I}V=-i[\hat{H}_w,\hat{H}_l]$.

As a final approximation, we assume that the superconducting phase is well-defined, i.e. that the work storage device is in a phase eigenstate (in the rotating frame)
\begin{equation}
\label{eq:rhotot}
\hat{\rho}_r=\hat{\rho}_S\otimes|\theta\rangle\langle\theta|,
\end{equation}
where the phase eigenstate is defined as
\begin{equation}
\label{eq:phaseeig}
|\theta\rangle = \frac{1}{\sqrt{2\pi}}\sum_{N=-\infty}^{\infty}{\rm e}^{-iN\theta}|N\rangle.
\end{equation}
In this case, we can replace the operator $\hat{\theta}$ with a real number in Eqs.~\eqref{eq:hint} (in the rotating frame) and in Eq.~\eqref{eq:curr}. The dynamics of the state of the system $\hat{\rho}_S$ is then determined by the Hamiltonian $\hat{H}_S+\hat{H}_{int}$.
This approximation is fully consistent with dropping the charging energy $E_C$. For large superconductors, the phase and not the total number of Cooper pairs is a well-defined quantity \cite{tinkham:book}. We note that in this case, the work storage device is in a superposition of all energy eigenstates [cf.~Eq.~\eqref{eq:phaseeig}]. The work can nevertheless be accessed via the measurable electrical current which determines the power. Finally, we note that the phase eigenstate provides a phase reference which is necessary to extract work from coherences as discussed in Refs.~\cite{aberg:2014,malabarba:2015,lostaglio:2015,korzekwa:2016,kwon:2018}.

We note that we disregard any cost associated to keeping time \cite{erker:2017,woods2016}.

%XXX I changed the style from bullet points to normal text (CB)
We conclude this section by summarizing the approximations we used:
% made on the Hamiltonian:
%\begin{enumerate}
%\item \label{ec}
(i) We dropped the charging energy: $E_C\rightarrow 0$. This is
justified for large superconductors as long as the charge accumulation
%XXX I replaced "imbalance" by accumlation since "charge imbalance" has
% a special meaning in nonequilibrium superconductivity 
does not become too large.
% \item \label{rwa}
(ii) We used the rotating-wave approximation which is expected to be valid as long as $\Omega\gg E_J\lambda$ and we are only interested in low-frequency observables.
% \item \label{phase}
(iii) We assumed the superconductor to be in a phase eigenstate, which
is consistent with (i).
%This approximation is completely consistent with approximation \ref{ec}.
%\end{enumerate}
Approximations (i) and (iii) 
% \ref{ec} and \ref{phase}
result in the standard Hamiltonian used for describing a Josephson junction coupled to a microwave resonator [given by $\Omega\hat{a}^\dagger\hat{a}-E_J\cos\left(2eVt+\hat{\varphi}\right)$] \cite{armour:2013,gramich:2013}.

\section{Fundamental limits}
\label{sec:limits}
{In this section, we briefly review previous results on the fundamental limits of extracting work from quantum states. To obtain such limits, one has to specify the control that one allows for. Here we are interested in the bounds obtained in {the abstract theory of quantum thermodynamics,} assuming full control in the sense that any unitary on the system can in principle be implemented. In this case, the maximum amount of work that can be extracted from a quantum state can be defined as the maximum by which its energy can be lowered using a unitary transformation.
The maximal amount of extractable work is then given by \cite{allahverdyan:2004,alicki:2013,perarnau:2015}}
\begin{equation}
\label{eq:wmaxun}
\begin{aligned}
W_{\rm max}(\hat{\rho})&=\underset{U}{\rm max}\:{\rm Tr}\{\hat{H}(\hat{\rho}-\hat{U}\hat{\rho}\hat{U}^\dagger)\}\\&={\rm Tr}\{\hat{H}\hat{\rho}\}-{\rm Tr}\{\hat{H}\hat{\pi}\},
\end{aligned}
\end{equation}
where the maximization is over all unitaries. We note that obtaining this limit in general not only requires the implementation of any possible unitary but also the conversion of the extracted energy into some useful form, justifying the notion of work. In the second line of the last expression, the state $\hat{\pi}$ denotes the passive state corresponding to $\hat{\rho}$, which is the state with lowest energy that can be obtained from $\hat{\rho}$ through a unitary transformation \cite{pusz:1978,lenard:1978}. If the eigenvalues of $\hat{\rho}$ are denoted as $p_i$, ordered in descending magnitude $p_{i+1}\leq p_i$, the passive state reads
\begin{equation}
\label{eq:passive}
\hat{\pi}=\sum_i p_i |E_i\rangle\langle E_i|,
\end{equation}
where $|E_i\rangle$ denote the eigenstates of $\hat{H}$, ordered in increasing energy $E_{i+1}\geq E_i$. We note that since $\hat{\rho}$ has the same entropy as the corresponding passive state, Eq.~\eqref{eq:wmaxun} can also be expressed as the difference of the free energies of the initial and the final state.

As an immediate consequence of Eq.~\eqref{eq:wmaxun}, one finds that all the energy stored in pure states can in principle be extracted as work since any pure state can be rotated into the ground state of the Hamiltonian. {This reflects the assumption of full control: all energy from a pure state can be extracted in the form of work as the state of the system is exactly known and all its degrees of freedom are accessible.}  
While considering a specific experiment severely restricts the unitary transformations that can be implemented, we find that the bound in Eq.~\eqref{eq:wmaxun} can be reached in our system for all Gaussian and Fock states. 

We note that just like in the resource theory of quantum thermodynamics \cite{janzing:2000,horodecki:2003,brandao:2013,horodecki:2013,skrzypczyk:2014,gour:2015,cwiklinski:2015,lostaglio:2015,lostaglio:2015prx,goold:2016,sparaciari:2017pra},
our setup is restricted to energy-conserving unitaries between the system and the work storage device {(note however the absence of a thermal bath). Since the work storage device remains in a phase eigenstate at all times, these unitaries act as energy non-conserving unitaries on the system alone. This makes Eq.~\eqref{eq:wmaxun} the relevant bound, even though it is not usually considered in the resource theory of quantum thermodynamics because the maximization might include unitaries that can not be implemented in an energy conserving way with the resources at hand. Under the restriction of energy conserving unitaries, work stored in coherences can only be extracted by making use of an additional source of coherence that acts as a phase reference \cite{aberg:2014,malabarba:2015,lostaglio:2015,korzekwa:2016,kwon:2018}.} In our case, this is provided by the work-storage device which can be approximated as being in a phase eigenstate. Alternatively, one could relax the unitaries to only preserve energy on average to access the work stored in the coherences \cite{skrzypczyk:2014,aberg:2014}.

Finally, we note that in the presence of a thermal bath, work can be extracted from passive states as long as their free energy is higher than that of the corresponding thermal state. The maximal amount of work that can be extracted from a state $\hat{\rho}$ is then still given by the free energy difference of the initial and final state, with the final state being a thermal state \cite{skrzypczyk:2014,aberg:2014}. {Making such beneficial use of the thermal bath usually requires a level of control over the coupling between system and bath that is not present in the experimental situation considered here.} {With the exception of thermal states, work can also be extracted from passive states when having access to multiple copies \cite{pusz:1978,lenard:1978,alicki:2013} or a catalyst \cite{sparaciari:2017}.}

\section{Extracting work from quantum states}
\label{sec:results}
In this section, we analyze the performance of the proposed engine. We will first consider the case where work is
extracted from a single quantum state. This scenario allows us to
compare the performance of the engine with respect to the fundamental bound given in
Eq.~\eqref{eq:wmaxun}. {In particular, we will see that for natural classes of quantum states (Gaussian states and Fock states), the bound can be saturated.} However, this assumes that the target state (the fuel) can
be prepared in the engine with a high fidelity and that the engine
can be switched on and off at will. To relax this demanding degree of control,
we investigate a continuous mode of operation, where a target state is being stabilized within the engine while work is being extracted. In this case, there is a steady influx of energy into the engine and the figure of merit will be the power produced by the engine, not the total work extracted from a single state. This scenario is relevant when the engine is used to extract work from a non-passive state that is continuously being generated by another machine. The continuous mode of operation allows us to consider situations that can be implemented with current-day technology \cite{hofheinz:2011,westig:2017,jebari:2018}. In particular, we discuss how our engine can be used to convert the power from a laser into an electrical current that flows against a voltage bias.

\begin{figure*}
\centering
\includegraphics[width=\textwidth]{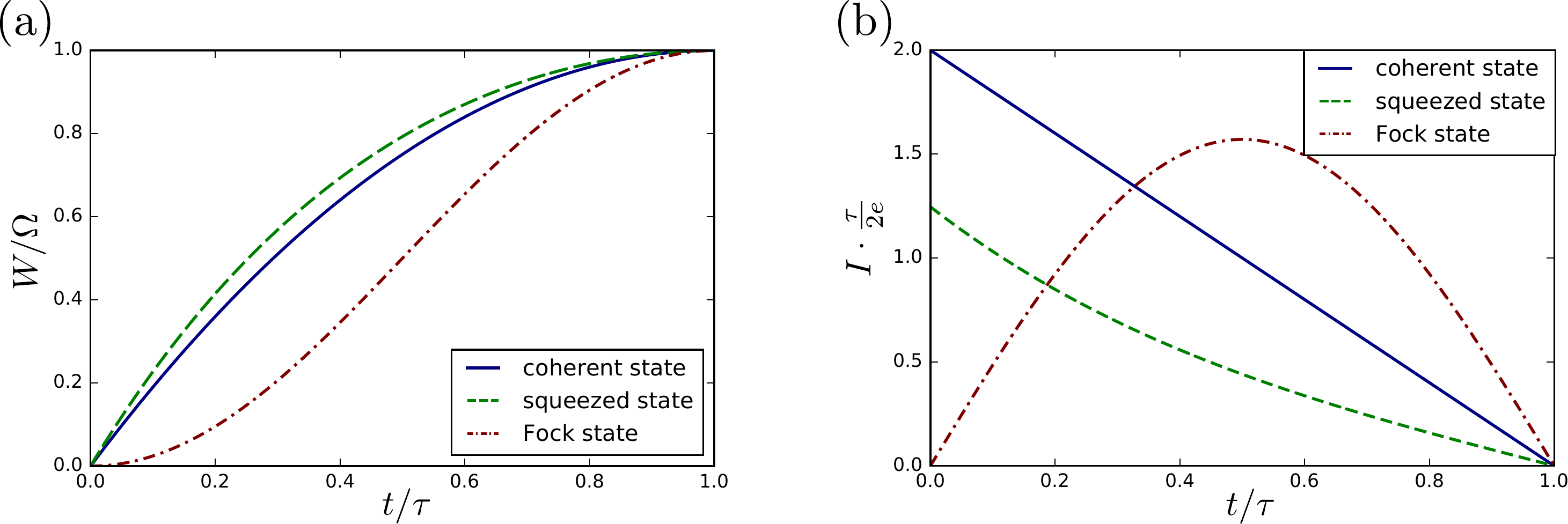}
\caption{Extracting work from quantum states. ${\rm (a)}$ Work
  extracted as a function of time. For all examples, all the energy
  (which is equal to $\Omega$, corresponding to one photon) is
  extracted after a time $\tau$ [which depends on the state, cf. the
  discussion below Eq.~\eqref{eq:uk} and Eq.~\eqref{eq:taun}]. For
  coherent and squeezed states, the energy stored in the off-diagonal
  density matrix elements is successfully converted into work. We note
  that coherence is crucial, as no work can be extracted from the
  completely decohered coherent state with one photon on average (the
  state is passive). We further find that coherence can speed up the
  work extraction, i.e., the work grows linearly in time at short times
  for the coherent and squeezed state whereas it grows quadratically
  in time for the Fock state. ${\rm (b)}$ Current as a function of
  time. The current is determined by the time derivative of the work. For states exhibiting coherence, the current is finite at $t=0$, underlining the coherence-induced speed-up of work extraction. The area underneath the curves gives the work divided by the voltage. Note that the voltage for the squeezed state is chosen as twice the value for the other states.}
  \label{fig:transient}
\end{figure*}

\subsection{Single quantum state}
To extract work from a single quantum state, we consider the situation where a state $\hat{\rho}_0$ is prepared in the resonator at $t=0$. We then unitarily time-evolve the state with the Hamiltonian given in Eq.~\eqref{eq:hint} (with the phase operator replaced by a real number). Since all the energy extracted from the state is turned into work, the work extracted at time $t$ is given by
\begin{equation}
\label{eq:work}
W=\Omega\left(\langle \hat{a}^\dagger\hat{a}\rangle_0-\langle \hat{a}^\dagger\hat{a}\rangle_t\right),
\end{equation}
where $\langle\cdots\rangle_t={\rm Tr}\{\cdots\hat{\rho}_t\}$ denotes the average at time $t$. We note that we recover for the power
\begin{equation}
\label{eq:powerss}
\langle \hat{P}\rangle_t=\partial_t W=V\langle \hat{I}\rangle_t.
\end{equation}
{We stress that Eq.~\eqref{eq:work} is only valid if we can neglect any dissipation to the environment, cf.~Sec.~\ref{sec:cont}.}

\subsubsection{Gaussian states}
\label{sec:gauss}
Any (single-mode) Gaussian state can be written as \cite{olivares:2012}
\begin{equation}
\label{eq:gaussstate}
\hat{\rho}_G=\hat{D}(\alpha)\hat{S}(\zeta)\hat{\rho}_\beta\hat{S}^\dagger(\zeta)\hat{D}^\dagger(\alpha),
\end{equation}
where the thermal state is given by
\begin{equation}
\label{eq:rhobeta}
\hat{\rho}_\beta=\frac{{\rm e}^{-\beta \Omega \hat{a}^\dagger\hat{a}}}{\mathcal{Z}},\hspace{.5cm} \mathcal{Z}={\rm Tr}\left\{{\rm e}^{-\beta \Omega \hat{a}^\dagger\hat{a}}\right\},
\end{equation}
and the displacement and squeeze operators read
\begin{equation}
\label{eq:displ}
\hat{D}(\alpha)={\rm e}^{\alpha\hat{a}^\dagger-\alpha^*\hat{a}},\hspace{.5cm}\hat{S}(\zeta)={\rm e}^{\frac{1}{2}[\zeta^*\hat{a}^2-\zeta (\hat{a}^\dagger)^2]}.
\end{equation}
These are unitary operators which satisfy $\hat{D}^\dagger(\alpha)=\hat{D}(-\alpha)$, $\hat{S}^\dagger(\zeta)=\hat{S}(-\zeta)$ and act on the annihilation operator as
\begin{equation}
\label{eq:actds}
\begin{aligned}
&\hat{D}^\dagger(\alpha)\hat{a}\hat{D}(\alpha)=\hat{a}+\alpha,\\&
\hat{S}^\dagger(\zeta)\hat{a}\hat{S}(\zeta)=\hat{a}\cosh(|\zeta|)-{\rm e}^{i\varphi}\hat{a}^\dagger\sinh(|\zeta|),
\end{aligned}
\end{equation}
where $\zeta=|\zeta|\exp(i\varphi)$. The passive state corresponding to the Gaussian state $\hat{\rho}_G$ is given by the thermal state $\hat{\rho}_\beta$.
The maximal amount of work can thus be extracted by transforming $\hat{\rho}_G$ into $\hat{\rho}_\beta$, undoing the displacement and the squeezing. From Eqs.~\eqref{eq:wmaxun} and \eqref{eq:actds}, we find
\begin{equation}
\label{eq:wmaxgauss}
W_{\rm max}(\hat{\rho}_G)=\Omega|\alpha|^2+\Omega\sinh^2(|\zeta|)\left(2n_\beta+1\right),
\end{equation}
with the thermal population
\begin{equation}
\label{eq:nbose}
n_\beta={\rm Tr}\{\hat{a}^\dagger\hat{a}\hat{\rho}_\beta\}=\frac{1}{{\rm e}^{\beta\Omega}-1}.
\end{equation}

As we will now show, our engine can extract the maximal amount of work from the state given in Eq.~\eqref{eq:gaussstate}.
To see this, we consider the limit $\lambda\ll1$ {(corresponding to impedances much smaller than the resistance quantum $h/e^2$)}. In this limit, the operators defined in Eq.~\eqref{eq:aops} become
\begin{equation}
\label{eq:aopsl0}
\hat{A}(k)\xrightarrow{\lambda\ll 1}\frac{(2\lambda)^k}{k!},
\end{equation}
where we suppressed the identity operator on the right-hand side.
This implies that by setting the voltage to $2eV=k\Omega$, the engine implements the Hamiltonian
\begin{equation}
\label{eq:haml0}
\hat{H}_k = -(2\lambda)^k\frac{E_J}{2k!}\left[i^k\hat{a}^k{\rm e}^{i\theta}+(-i)^k(\hat{a}^\dagger)^k{\rm e}^{-i\theta}\right].
\end{equation}
For $k=1,2$ this Hamiltonian results in the time-evolution operators
\begin{equation}
\label{eq:uk}
\hat{U}_k(t)={\rm e}^{-i\hat{H}_kt}=\begin{cases}
\hat{D}\left(\lambda E_Jt{\rm e}^{-i\theta}\right)\,\,\,\,\,\,\,\,{\rm for}\,\,k=1,\\
\hat{S}\left(2i\lambda^2E_Jt{\rm e}^{-i\theta}\right)\,\,{\rm for}\,\,k=2.
\end{cases}
\end{equation}
The maximal amount of work can then be extracted from a Gaussian state by first displacing it to the origin followed by undoing the squeezing. For the displacement, we set $2eV=\Omega$ (i.e., $k=1$) and the superconducting phase difference to $\theta=\pi-\arg(\alpha)$ and evolve the system for the time $\tau_\alpha=|\alpha|/(\lambda E_J)$. The time-evolution operator then reads $\hat{D}(-\alpha)=\hat{D}^{-1}(\alpha)$ which cancels the displacement in Eq.~\eqref{eq:gaussstate}. To extract the remaining work from the resulting squeezed state, the voltage is changed to $2eV=2\Omega$ and the phase difference to $\theta=-\pi/2-\arg(\zeta)$ and the system is evolved for the time $\tau_\zeta = |\zeta|/(2\lambda^2 E_J)$. The corresponding time-evolution operator is given by $\hat{S}(-\zeta)=\hat{S}^{-1}(\zeta)$, undoing the squeezing in Eq.~\eqref{eq:gaussstate} resulting in the passive state $\hat{\rho}_\beta$.

\begin{figure}
\centering
\includegraphics[width=\columnwidth]{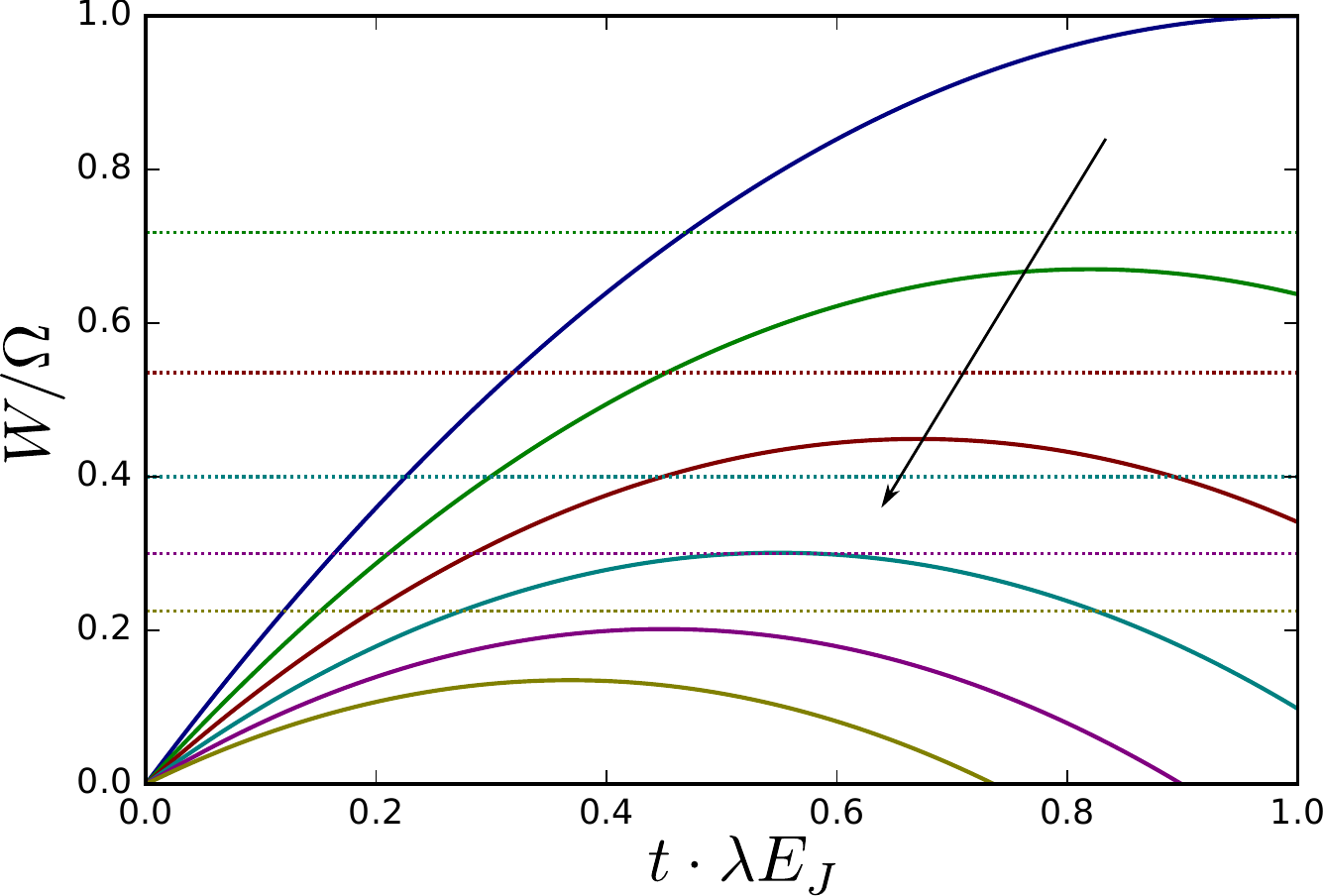}
\caption{Effect of decoherence. The different curves show the work extracted from a coherent state subject to different amounts of decoherence. The dotted lines show the corresponding maximal amount of extractable work [cf.~Eq.~\eqref{eq:wmaxun}]. Decoherence is taken into account by exponentially suppressing the off-diagonal density matrix elements by $\exp(-\gamma(n-n')^2)$, where $n$ and $n'$ are the respective eigenvalues of the number operator. For increasing $\gamma$, the amount of work extracted is reduced. Furthermore, the maximal amount of extractable work can no longer be reached. This implies that not all extractable work can be extracted by displacing the state. We note that the fully decohered state with mean energy $\Omega$ is a passive state. Values of $\gamma$ are $[0,0.4,0.8,1.2,1.6,2]$ and increase in the direction of the arrow.}
  \label{fig:decoherence}
\end{figure}

The extraction of work from a coherent state [Eq.~\eqref{eq:gaussstate} with $\beta\rightarrow\infty$ and $\zeta=0$] as well as a squeezed vacuum [Eq.~\eqref{eq:gaussstate} with $\beta\rightarrow\infty$ and $\alpha=0$] is illustrated in Fig.~\ref{fig:transient}. Since these states are pure, all their energy can be extracted as work. We stress that the coherences between energy eigenstates are crucial for the work extraction procedure. Indeed, for a coherent state with mean energy $\Omega$ (i.e. including a single photon on average), the decohered state obtained by setting the off-diagonal density matrix elements to zero is a passive state. The effect of decoherence on the work extraction is illustrated in Fig.~\ref{fig:decoherence}. There we consider a coherent state which is subject to decoherence before the engine is switched on. To this end, we suppress the off-diagonal density elements by a factor $\exp(-\gamma(n-n')^2)$, where $n$ and $n'$ are the respective eigenvalues of the number operator. This corresponds to time-evolving the state by the Markovian master equation
\begin{equation}
\label{eq:decoherence}
\partial_t\hat{\rho}=2\Gamma\hat{n}\hat{\rho}\hat{n}-\Gamma\left\{\hat{n}^2,\hat{\rho}\right\}
\end{equation}
for a time $t=\gamma/\Gamma$ before the engine is switched on and the time evolution is determined by the Hamiltonian alone. In the last expression, $\hat{n}=\hat{a}^\dagger\hat{a}$ and $\{.,.\}$ denotes the anti-commutator.

As illustrated in Fig.~\ref{fig:decoherence}, decoherence reduces the maximal amount of work that can be extracted from the state. Furthermore, the engine is no longer able to extract the maximal amount of work since {displacing the state no longer results in a passive state, i.e., the unitary which maximizes Eq.~\eqref{eq:wmaxun} is no longer given by a displacement operator}.

As shown in Eq.~\eqref{eq:uk}, any (single-mode) Gaussian unitary can be implemented by the engine. This implies that, if the engine is restricted to small $\lambda$, the amount of work the engine can extract is given by the energy difference of the state and the corresponding Gaussian passive state \cite{brown:2016}.

\begin{figure*}
\centering
\includegraphics[width=\textwidth]{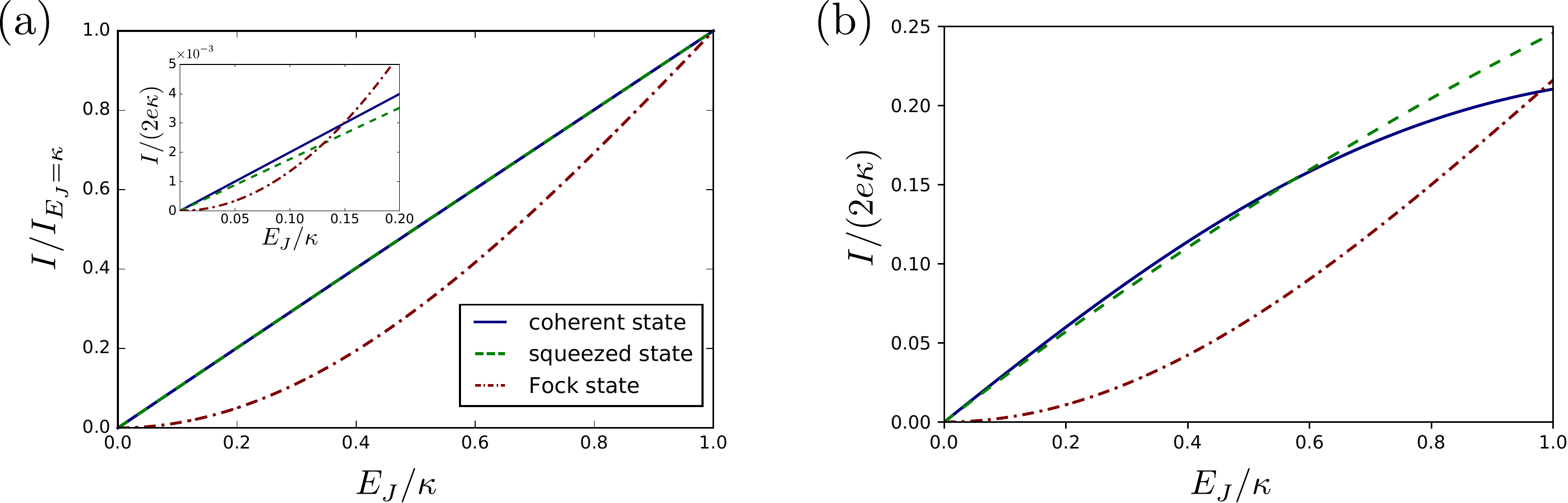}
\caption{Extracting work from continuously stabilized quantum
  states. ${\rm (a)}$ Current as a function of $E_J$, normalized by
  its value at $E_J=\kappa$. The values of $\lambda$ are chosen such
  that the engine can extract all the energy of a single state
  (i.e. for coherent and squeezed states $\lambda=0.01\ll 1$, for Fock
  state $\lambda=1/\sqrt{2}$). All states have mean energy $\Omega$
  and we chose the optimal $\theta$ for work extraction (i.e.,
  $\alpha=-\exp(-i\theta)$ for the coherent state, $\zeta=-i\exp(-i\theta)\sinh^{-1}(1)$ for the
  squeezed state, and $n=1$ for the Fock state). The inset shows the
  low $E_J$ behavior where states that exhibit coherence show a linear
  scaling, outperforming the Fock state which shows a quadratic
  scaling. The current in the inset is not normalized.
%XXX why? PH: To see which state performs better on an absolute scale (i.e. in the same engine)
  In this regime, the numerical solutions can be perfectly reproduced by
  the analytical expressions given in \eqref{eq:resetcurrcoh} -- \eqref{eq:resetcurrfockxn}.
  ${\rm (b)}$ Same as ${\rm (a)}$ but with $\lambda=1/\sqrt{2}$ for
  all states; all currents are given in units of $2e\kappa$. To obtain
  the power, the currents have to be multiplied by $V$ (i.e., $\Omega$
  for the coherent state and Fock state, $2\Omega$ for the squeezed state).}
  \label{fig:continuous}
\end{figure*}

\subsubsection{Fock states}
Since Fock states are pure states, all their energy can in principle be extracted as work by unitarily transforming them into the ground state. However, because their Wigner functions are rotationally invariant, no energy can be extracted by applying displacement and squeeze operators. To extract energy from Fock states, we thus have to make use of the non-linearity that is provided by the Josephson junction [cf.~Eq.~\eqref{eq:hamint}], {which allows for implementing non-Gaussian time-evolution operators}. To show that our engine is able to extract all the energy stored in Fock states, we focus on the state which includes $n$ photons $|n\rangle$.
We then set the external voltage to $2eV=n\Omega$, such that the Hamiltonian is given by Eq.~\eqref{eq:hint}, with $k=n$. In this case, the Hamiltonian connects the states
\begin{equation}
\label{eq:connections}
|0\rangle\leftrightarrow|n\rangle\leftrightarrow|2n\rangle\leftrightarrow \cdots.
\end{equation}
We then set $\lambda$ to fulfill
\begin{equation}
\label{eq:laguerrezero}
4\lambda^2=x_{n},\hspace{1cm}L_{n}^{(n)}(x_{n})=0,
\end{equation}
where $x_{n}$ denotes a zero of the generalized Laguerre
polynomial. This cuts the second arrow
% connection
in Eq.~\eqref{eq:connections} and leads to an effective Hamiltonian describing the relevant subspace
\begin{equation}
\label{eq:hameff}
%\hat{H}_{\rm eff}=-\frac{E_J}{2}\sqrt{\frac{(x_{m})^{m}}{m!}}{\rm e}^{-\frac{x_{m}}{2}}{\rm e}^{i\theta}i^{m}|0\rangle\langle m|+H.c.
\hat{H}_{\rm eff}=-\frac{\pi}{2\tau_n}{\rm e}^{i\theta}i^{n}|0\rangle\langle n|+H.c.,
\end{equation}
where 
\begin{equation}
\label{eq:taun}
\tau_{n}= \frac{\pi}{E_J}\sqrt{\frac{n!}{(x_{n})^{n}}}{\rm e}^{x_{n}/2}.
\end{equation}
The time evolution will lead to Rabi oscillations between the states
$|n\rangle$ and $|0\rangle$.  All the energy in the Fock state can
thus be converted into electrical work by evolving for a time
$\tau_{n}$.  This is illustrated in Fig.~\ref{fig:transient}. Note that the phase $\theta$ is irrelevant in this case due to the rotational symmetry of the Wigner function of Fock states.

While the extracted work starts out linearly as a function of time for the coherent and the squeezed state, it grows quadratically for the Fock state. This is a consequence of the fact that the Fock state does not exhibit any coherence in the energy eigenbasis. To extract work from a diagonal state, populations from higher energies have to be moved to lower energies. Under unitary time-evolution, populations are exchanged by first turning them into coherences between energy eigenstates which are then turned into populations again. This results in the quadratic behavior of the extracted work at short times. For states that already exhibit coherences at $t=0$, these coherences can directly be turned into populations resulting in a linear growth of the extracted work. Coherences thus enable a speed-up of work extraction at small times. As we will see below, this effect is also manifest in the continuous operation of the engine, allowing for increased output powers in the steady state.

{We note that by operating the machine in reverse, Fock states can be prepared from the vacuum and Gaussian states can be prepared from thermal states.}

\subsubsection{Multi-mode states}
In this section, we extend our results to multi-mode states, where we show that the maximum amount of work can be extracted from all Gaussian and Fock states. To this end, we consider multiple resonators that are coupled to the Josephson junction. In this case, the phase $\hat{\varphi}$ in Eq.~\eqref{eq:hamtot} has to be replaced by a sum over all resonators $\sum_j\hat{\varphi}_j$. By tuning the voltage, different terms can be tuned into resonance just like for a single resonator. Considering incommensurate frequencies $\Omega_j$, local Hamiltonians of the form \eqref{eq:hint} can still be implemented by setting the voltage equal to $2eV=k\Omega_j$. Furthermore, in the low-$\lambda$ limit, a beam-splitter Hamiltonian between modes $l$ and $j$ can be implemented by setting the voltage equal to $2eV=\Omega_l-\Omega_j$ \cite{hofer:2016prb}. Similarly, two-mode squeezing can be obtained by choosing $2eV=\Omega_l+\Omega_j$ \cite{westig:2017}. Therefore, all Gaussian unitaries can be implemented in the multi-mode case as well. The maximal amount of work can thus be extracted from all multi-mode Gaussian states.

The multi-mode case is of particular interest since it was recently shown that a quantum advantage can result in an enhanced power output in this scenario \cite{hovhannisyan:2013,binder:2015njp,campaioli:2017,ferraro:2018}. To see this, consider work extraction from the state $|n\rangle^{\otimes M}$. Using $M$ engines, each coupled to a single resonator, the maximal work $n\sum_j\Omega_j$ can be extracted in the time $\tau_n$ [cf.~Eq.~\eqref{eq:taun}]. By directly coupling the state $|n\rangle^{\otimes M}$ to the vacuum, one can possibly extract the same amount of work in a faster time since the state takes a shorter route through Hilbert space. In our engine, setting the voltage equal to $2eV=n\sum_j\Omega_j$ couples the states $|0\rangle^{\otimes M}\leftrightarrow|n\rangle^{\otimes M}\leftrightarrow|2n\rangle^{\otimes M}\leftrightarrow\cdots$ (for incommensurate frequencies). By setting a single parameter $\lambda_0$ equal to Eq.~\eqref{eq:laguerrezero}, the transition from $|n\rangle^{\otimes M}\rightarrow|2n\rangle^{\otimes M}$ can be suppressed, just like for the single-mode Fock state. We then obtain the effective Hamiltonian
\begin{equation}
\label{eq:hameffmul}
\hat{H}_{\rm eff}=-\frac{\pi}{2\tau_{n,M}}{\rm e}^{i\theta}i^{n\cdot M}|0\rangle\langle n|^{\otimes M}+H.c.
\end{equation}
This Hamiltonian extracts the maximal work in the time $\tau_{n,M}$. Minimizing $\tau_{n,M}$ with respect to all $\lambda_{j\neq 0}$ except the one that was fixed to obtain the effective Hamiltonian results in
\begin{equation}
\label{eq:taunm}
\tau_{n,M}=\tau_n\left(\sqrt{\frac{n!}{n^n}}{\rm e}^{\frac{n}{2}}\right)^{M-1}>\tau_n,
\end{equation}
where the last inequality holds for $M>1$. It is therefore always beneficial to use a single engine for each resonator. 

However, designing multiple engines might be more difficult than designing a single one. One can therefore ask if the global Hamiltonian in Eq.~\eqref{eq:hameffmul} results in a speed-up compared to using a single engine to extract work from the resonators one after the other. Adressing the resonators one by one allows for extracting the maximal work after the time $M\tau_n$. One can show that (for $M>1$),
\begin{equation}
\label{eq:taumnsb}
\tau_{n,M}>M\tau_n \hspace{1cm}{\rm for}\hspace{.5cm}n+M>4.
\end{equation}
We thus find that for a large number of photons involved, it is always beneficial to extract the photons one by one, using only local Hamiltonians. The reason for this is that as the photon number increases, it becomes increasingly unlikely that a Cooper pair absorbs all photons at once. Therefore it becomes beneficial to extract the work in multiple steps, where a Cooper pair absorbs a smaller number of photons in each step.
We note that the results for the state $|n\rangle^{\otimes M}$ can straightforwardly be generalized to arbitrary Fock states.

\subsection{Continuous operation}
\label{sec:cont}
In this section we analyze the engine in {\it continuous operation},
where a particular quantum state $\hat{\rho}_0$ is stabilized. {In this case, the engine becomes autonomous as no time-dependent external control is needed.} For pure states we will use the notation $\hat{\rho}_0=\ket \psi \bra \psi$. In analogy to the previous section on extracting work from a particular quantum state, we study the cases where
$\ket \psi$ is a coherent state $\ket \alpha$, a Fock state $\ket n$, or a squeezed state $\ket{0,\zeta}$. {We note that all these states can be stabilized in present-day experiments \cite{poyatos:1996,breitenbach:1997,sayrin:2011,holland:2015,souquet:2016}}. For each state, we will focus on the parameters that allow for maximal work extraction in the scenario discussed in the previous section.

{We will consider two models to describe the engine: a heuristic reset model, which works for any state $\hat{\rho}_0$, and a microscopic Lindblad model which describes the stabilization of a coherent state using a coherent drive in the presence of dissipation. We find good agreement between the reset model and the Lindblad model, indicating that the reset model is well suited to obtain a qualitative picture of the work extraction process.} {In both cases, the time-evolution is no longer governed by the Hamiltonian alone. This opens a dissipative channel for energy, such that Eq.~\eqref{eq:work} is no longer valid and energy stored in the quantum state can irretrievably be lost to the environment.}

%XXX_NL: After discussion: Numerically do only n=1, Analytically Patrick might look into n unequal 1.

%XXX the squeezing parameter here is taken from Niels' numerical calculation and is also consistent with the convention introduced just above Equation (66) in ''notes_patrick.pdf''. However, it comes with a different sign as compared to the caption of Fig. 4. PH: Caption in Fig. 4 seems to be correct and consistent with 'notes_patrick.pdf' (negative r results in positive current in Eq. (66))

\subsubsection{Reset model}

We model all cases with a reset model of the form \cite{linden:2010prl,barra:2017}
\begin{align}
\label{eq:reset}
&\partial_t\hat{\rho} = -i[\hat{H}_{int},\hat{\rho}] + \kappa \left(\hat{\rho}_0 - \hat{\rho} \right)
\end{align}
where $\hat{H}_{int}$ is the respective work extracting Hamiltonian and the non-unitary term stabilizes $\hat{\rho}_0$ at a dissipation rate $\kappa$.
As before, the Hamiltonian and current are given by
Eq.~\eqref{eq:hint} and Eq.~\eqref{eq:curr}, with $\hat{\theta}$ replaced by a real number.

To obtain analytical results for the case of a {\it coherent state}, we use the $\lambda\ll1$
approximation  given in Eq.~\eqref{eq:aopsl0} and we set $2eV=\Omega$. The resulting approximate Hamiltonian is given in Eq.~\eqref{eq:haml0} with $k=1$. Within this approximation, the current can be calculated analytically from the reset model. In the steady state, we find
\begin{equation}
\label{eq:resetcurrcoh}
I_\alpha=4e \lambda E_J \left( |\alpha| -\lambda \frac{E_J}\kappa \right),
\end{equation}
where we chose the optimal phase for work extraction, i.e. $\theta=\pi-\arg(\alpha)$ [cf.~Sec.~\ref{sec:gauss}].

For the {\it squeezed state}, we again make the small-$\lambda$ approximation on the Hamiltonian, but we set $2eV=2\Omega$ (i.e., $k=2$), such that the work extracting Hamiltonian generates a squeeze operator. We then find the current 
\begin{equation}
\label{eq:resetcurrsq}
I_\zeta=2e \lambda^2 E_J \frac{\sinh \left(2|\zeta|\right) +4\lambda^2 \left(E_J/\kappa \right) \cosh \left(2|\zeta|\right)}{1+16 \lambda^4 \left(E_J/\kappa\right)^2},
\end{equation}
where the optimal phase $\theta=-\pi/2-\arg(\zeta)$ is chosen.

For a {\it Fock} state with $n$ photons, we use the full, non-linear Hamiltonian in Eq.~\eqref{eq:hint} with $k=n$ (corresponding to $2eV=n\Omega$). There are therefore no restrictions on $\lambda$. However, to make analytical progress, we treat the unitary term in Eq.~\eqref{eq:reset} as a perturbation and calculate the current to lowest order in $E_J/\kappa$.  Using perturbation theory for Lindblad operators \cite{Li2014} we obtain the current
\begin{equation}
\label{eq:resetcurrfock}
I_n=e\frac{E_J^2}{\kappa}(4\lambda^2)^{n}{\rm e}^{-4\lambda^2}\left\{\frac{1}{n!}-\frac{n!}{(2n)!}\left[L_n^{(n)}(4\lambda^2)\right]\right\}.
\end{equation}
%For the Fock state including a single photon, this reduces to
%\begin{equation}
%\label{eq:resetcurrfock1}
%I_{n=1}=e \frac{E_J^2}\kappa 4\lambda^2 {\rm e}^{-4\lambda^2} \left(1-2\left(1-2\lambda^2\right)^2\right) .
%\end{equation}

For the special value of $\lambda$ given in Eq.~\eqref{eq:laguerrezero}, the master equation in Eq.~\eqref{eq:reset} only couples the states $|0\rangle$ and $|n\rangle$. In this case, the reset model can be solved analytically resulting in the current
\begin{equation}
\label{eq:resetcurrfockxn}
I_n = \frac{e\kappa E_J^2(x_n)^n{\rm e}^{-x_n}}{\kappa^2 n!+E_J^2(x_n)^n{\rm e}^{-x_n}},
\end{equation}
%For $n=1$ (with $x_1=2$) this reduces to
%\begin{equation}
%I_{n=1}=2e \frac{\kappa E_J^2 \exp[-2]}{\kappa^2+2E_J^2 \exp[-2]}\:.
%\label{curves3}
%\end{equation}
 which is valid for all $E_J$.
Note that for small $E_J$, the current of the Fock state scales with $E_J^2$, while
both coherent and squeezed state result in a scaling linear in $E_J$. Analogously to the last section, the linear scaling arises from the coherences in the energy eigenbasis which allow the undisturbed state to carry a current (i.e., ${\rm Tr}\{\hat{I}\hat{\rho}_0\}\neq 0$). 
For small $E_J$, coherence thus increases the power output of the engine.

The currents obtained when operating the engine continuously are illustrated in Fig.~\ref{fig:continuous}. The exact numerical solutions were obtained with QuTiP \cite{johansson:2013}.
Figure \ref{fig:scans_other} in App.~\ref{app:comparison} illustrates the regime of validity of the analytical expressions obtained in this section.

\subsubsection{Lindblad model}

\begin{figure}
\centering
\includegraphics[width=0.5\textwidth]{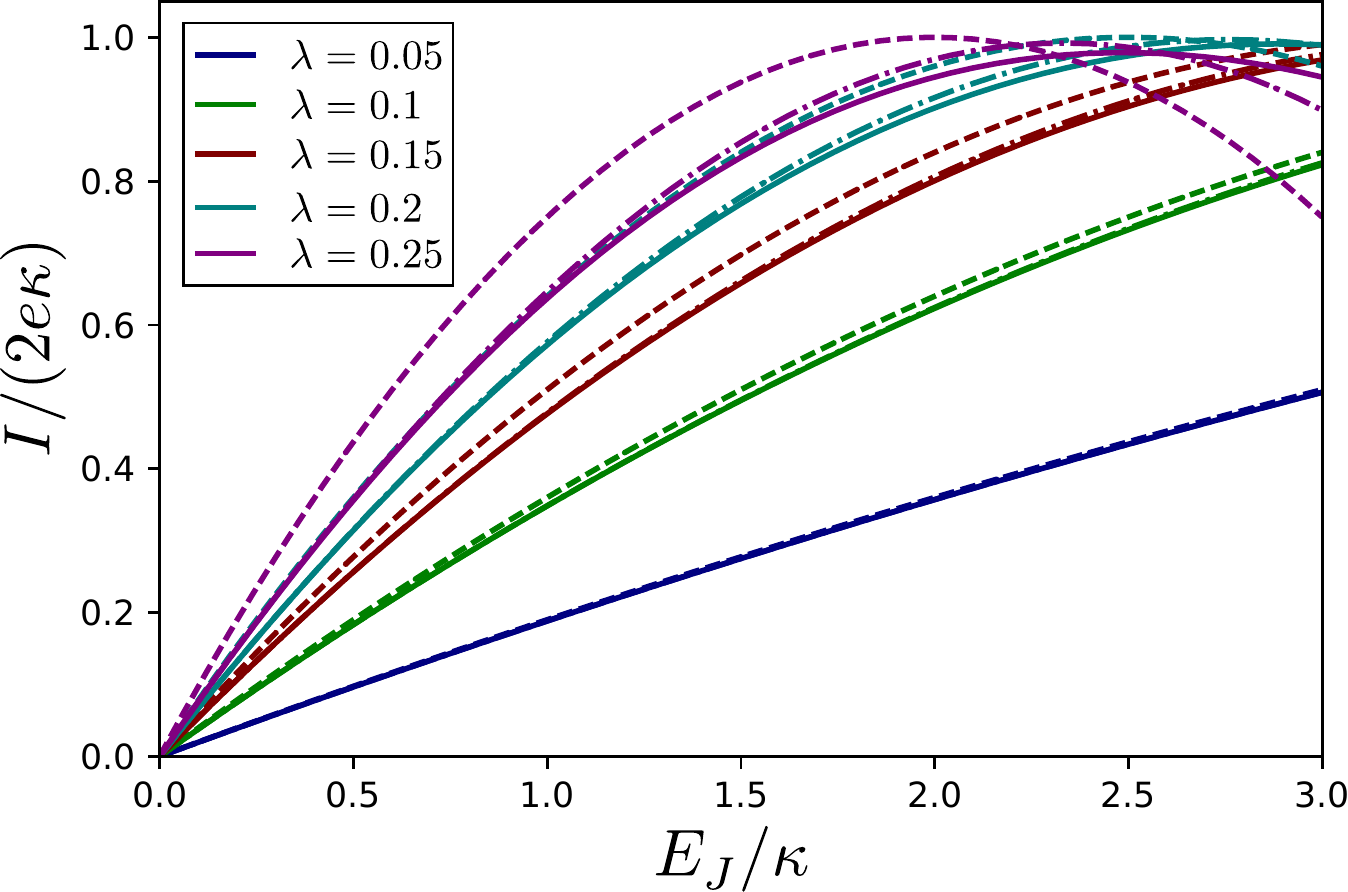}
\caption{{Comparison of the reset model and the Lindblad model for extracting work from a continuously stabilized coherent state.
The current $I$ is shown as a function of the Josephson energy $E_J$ for different values of the coupling parameter $\lambda$.} The full lines are solutions of the reset model in Eq.~\eqref{eq:reset}, the dashed lines are solutions of the Lindblad master equation given in Eq.~\eqref{eq:lindblad}. The dash-dotted lines are the analytical low-$\lambda$ expressions given in Eqs.~\eqref{eq:resetcurrcoh} and \eqref{eq:lindbladcurr} which hold both for the reset model as well as the Lindblad model. As $\lambda$ is increased, the low-$\lambda$ approximation breaks down and the reset model starts to show deviations from the Lindblad model. As expected, we find good agreement between the reset model and the Lindblad model as long as $\lambda E_J/\kappa$ is small.}
  \label{fig:scans_coherent}
\end{figure}

In this section, we consider the case where an external, resonant coherent drive (given, e.g., by a laser with frequency $\Omega$) drives the state in the resonator towards a displaced state. The Josephson Hamiltonian is then used to convert the work produced by the coherent drive into electrical work. The engine therefore converts one form of work into another one. This scenario is described by the master equation (in the rotating frame)
\begin{equation}
\label{eq:lindblad}
\begin{aligned}
\partial_t\hat{\rho}=&-i[\hat{H}_{d}+\hat{H}_{int},\hat{\rho}]\\&+2\kappa(n_\beta+1)\mathcal{D}[\hat{a}]\hat{\rho}+2\kappa n_\beta\mathcal{D}[\hat{a}^\dagger]\hat{\rho},
\end{aligned}
\end{equation}
where the coherent drive is captured by
\begin{equation}
\label{eq:drivenham}
\hat{H}_d=f\hat{a}^\dagger+f^*\hat{a},
\end{equation}
and the Josephson Hamiltonian $\hat{H}_{int}$ is given in Eq.~\eqref{eq:hint}. In addition to the unitary evolution, the resonator is coupled to a thermal bath at inverse temperature $\beta$ which leads to the dissipative terms in Eq.~\eqref{eq:lindblad} where
\begin{equation}
\label{eq:dissipator}
\mathcal{D}[\hat{A}] \hat{\rho} =  \hat A \hat{\rho} \hat A^\dagger -  \frac{1}{2}\{\hat A^\dagger\hat A, \hat{\rho}\} .
\end{equation}
Note that the validity of such a dissipation model was recently verified in a similar system \cite{hofer:2017njp,gonzalez:2017}. 
In the absence of $\hat{H}_{int}$, the laser drive together with the dissipation stabilizes the state
\begin{equation}
\label{eq:steadylindblad}
\hat{\rho}_0=\hat{D}(\alpha)\hat{\rho}_\beta \hat{D}^\dagger(\alpha),
\end{equation}
where $\alpha=-if/\kappa$. This state reduces to a pure (coherent) state for $\beta\rightarrow\infty$. For small $\hat{H}_{int}$ (i.e., small $\lambda E_J/\kappa$), where the steady state remains close to $\hat{\rho}_0$, we find quantitative agreement between the predictions of the Lindblad equation in Eq.~\eqref{eq:lindblad} and the reset model in Eq.~\eqref{eq:reset}, see Fig.~\ref{fig:scans_coherent}. The good agreement for small $\lambda$ and $E_J/\kappa$ shows the adequacy of using the heuristic reset model to describe physical systems in leading order. For larger $E_J/\kappa$ the reset model becomes extremely nonlinear, as all states decay directly to the stabilized state.

The power that enters the resonator from the coherent drive reads
\begin{equation}
\label{eq:pin}
P_{in}= -i\langle [\Omega\hat{a}^\dagger\hat{a},\hat{H}_d]\rangle=\Omega\langle -if\hat{a}^\dagger+if^*\hat{a}\rangle.
\end{equation}
We note that in the laboratory frame, where the driving Hamiltonian is time-dependent, this is equivalent to the standard definition $P_{in}=\langle\partial_t\hat{H}_d(t) \rangle$.

From Eq.~\eqref{eq:lindblad}, we find
\begin{equation}
\label{eq:systemen}
\partial_t\Omega\langle\hat{a}^\dagger\hat{a}\rangle=P_{in}-P-J,
\end{equation}
where $P=\langle \hat{I}\rangle V$ is the electrical power that is generated, and the heat current (leaving the system) is defined as
\begin{equation}
\label{eq:heatcurr}
J=2\kappa \Omega\left(\langle \hat{a}^\dagger\hat{a}\rangle-n_\beta\right).
\end{equation}
In the steady state, we find the first law of thermodynamics $P_{in}=P+J$. The power of the coherent drive is either converted into electrical power or it is dissipated to the environment.

We now consider the limit $\lambda\ll 1$, a voltage $2eV=\Omega$ [i.e., $k=1$ in Eq.~\eqref{eq:hint}] and a phase $\theta=-\arg(f)-\pi/2$, which is the optimal case for work extraction. We then find $\hat{H}_{int}=-(\lambda E_J/|f|)\hat{H}_d$. While the coherent drive displaces the state towards $|\alpha\rangle$, the Josephson interaction induces a displacement in the opposite direction, lowering the energy of the state which results in the generated electrical power. In this case, the steady state is still a displaced thermal state and we obtain for the produced current
\begin{equation}
\label{eq:lindbladcurr}
I=4e\lambda\frac{E_J}{\kappa}\left(|f|-\lambda E_J\right).
\end{equation}
Note that this expression is independent of temperature. Using $\alpha=-if/\kappa$, the last equation is equivalent to the current obtained from the reset model in Eq.~\eqref{eq:resetcurrcoh}. We note that the equivalence between the reset model and the Lindblad master equation only holds for small $\lambda$, as illustrated in Fig.~\ref{fig:scans_coherent}.

We can then ask how efficiently the power is transferred from the laser to the electronic current. The efficiency of this process is given by
\begin{equation}
\label{eq:efficiency}
\eta = \frac{P}{P_{in}}=\frac{P}{P+J}=\frac{\lambda E_J}{|f|}.
\end{equation}
In the steady state, the heat current has to be positive $J\geq0$ due to the second law of thermodynamics preventing the production of work using only heat from a single bath. Whenever $P, P_{in}>0$ (i.e., in the regime where power from the laser is converted into electrical power), the efficiency thus fulfills $\eta\leq 1$. Interestingly, this inequality can be saturated. However, this happens when the Hamiltonian of the drive exactly cancels the work extraction Hamiltonian (i.e., when $\hat{H}_d=-\hat{H}_{int}$). The system then remains in a thermal state at inverse temperature $\beta$, resulting in $P=P_{in}=0$.
We thus find that (in the steady state) we can only convert power from the laser to electrical power with unit efficiency at the point where the converted power vanishes. When converting energy at a finite rate, we necessarily have dissipation (i.e., $J>0$). This is reminiscent of heat engines, where the optimal efficiency (the Carnot efficiency) can only be obtained at the point of reversibility, where the produced power goes to zero. It is interesting that such a limitation seems to not only hold for converting heat into work, but also when extracting work from one physical degree of freedom and storing it in another degree of freedom.

%XXX will this section materialize? 
%XXX_NL: commented out for now
%\section{Imperfections}
%\label{sec:imperfections}
%\begin{itemize}
%\item Charging energy?
%\item Decoherence, noise, and leakage
%\item Numerical study
%\item Numerical results for large $\lambda$?
%\end{itemize}

\section{Conclusions}
\label{sec:conclusions}
To realize the full potential of quantum thermal machines as energy harvesters, it is crucial to understand the complete process of work extraction, starting from a non-equilibrium situation and ending with the performance of a useful task or with energy being stored in a controlled way (e.g. in a battery). Here we have considered the question of how to extract work stored in a resource quantum state and convert it into a more stable and controllable form, in this case an electrical current.

To this end, we have described and analyzed an engine that extracts energy from quantum states in the form of an electrical current against a voltage bias. The engine consists of a resonator, containing the quantum state {(the fuel of the engine)}, and a Josephson junction, which is used to extract work from the state by inelastic Cooper pair tunneling. The phase difference across the junction acts as a phase reference, allowing the extraction of work from coherences in the energy eigenbasis. We have demonstrated that the engine can extract the maximal amount of work from all Gaussian states and Fock states. We have further considered the scenario where work is extracted from a target state that is continuously being stabilized in the resonator. In both scenarios, we have found that coherence can enhance the extracted power. Finally, we have considered the case where the engine is used to convert the power provided by a laser into electrical power. We have found that this conversion necessarily leads to losses when it is performed at a finite rate, in analogy to heat engines which can only reach their maximal efficiency when the extracted power vanishes. {As our setup is realistic \cite{hofheinz:2011,westig:2017,jebari:2018} and saturates fundamental bounds, our proposal promises to open up interesting possibilities for near-future experiments.}

{Our work further rises some interesting questions that we leave open for future research. For instance, we only considered a thermal bath to model the dissipation usually present in experiments. In the resource theory of quantum thermodynamics, the presence of a thermal bath can result in better work extraction procedures. Finding experimental situations where those procedures can be implemented would further underline the experimental relevance of abstract resource theories and the role of thermal baths as resources. Another interesting direction is provided by going beyond the rotating wave approximation that allows us to consider an energy conserving unitary operation. This allows for the investigation of non-energy conserving interactions, providing the opportunity to investigate work extraction beyond the resource theory of quantum thermodynamics on a concrete system.
Finally,} we have only focused on mean values, {leaving the interesting question of fluctuating work open for future studies}. The investigation of fluctuations requires one to take into account the back action of the measurement on the quantum state. Such a measurement necessarily has to take place at some stage of the work extraction. {A key advantage of devices where work is provided by an electrical current is provided by the large body of literature that investigates the measurement of higher moments, finite frequency noise, and even the full probability distribution (full counting statistics) of the electrical current (cf.~\cite{nazarov:book2} and references therein).}

Making practical use of quantum heat engines requires a better understanding of the intricate interplay between work production, extraction, and measurement (and possibly feedback) in concrete physical systems. The present work constitutes a step in this direction and hopefully stimulates further activity on this intriguing subject.

\section*{Acknowledgments}
We thank M. Perarnau-Llobet and A. Roulet for valuable feedback. This work was financially supported by the Swiss National Science Foundation (Grants
200021\_169002 and 200020\_175596) and the NCCR
Quantum Science and Technology (QSIT). PPH acknowledges financial
support by the Swedish Research Council.

\appendix

\section{Comparison of numerical and analytical results}
\label{app:comparison}
\begin{figure*}
\includegraphics[width=\textwidth]{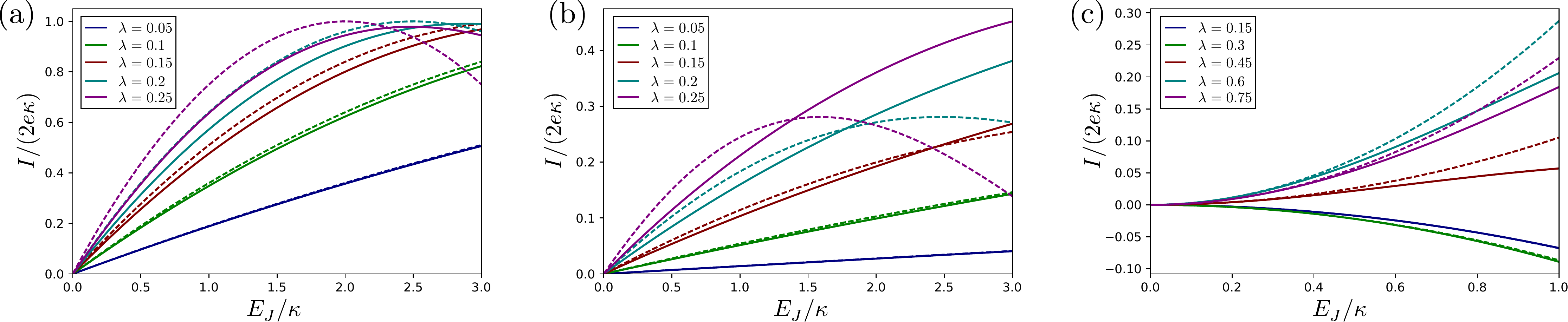}
\caption{Current as a function of  $E_J$ for the
  stabilization of (a) a coherent state, (b) a squeezed state, and (c) a Fock state. The solid lines are numerical results of the reset model given in Eq.~\eqref{eq:reset} and the dashed lines are analytical approximations given in Eqs.~(\ref{eq:resetcurrcoh}-\ref{eq:resetcurrfock}). For the coherent and the squeezed state, the analytical solution is exact in the small $\lambda$ limit, for the Fock state it is exact in the limit $E_J/\kappa\ll 1$.}
  \label{fig:scans_other}
\end{figure*}

Figure \ref{fig:scans_other} provides a comparison of our analytical approximations and numerical results for the reset model discussed in Sec.~\ref{sec:cont}. We see that agreement is excellent in the regimes where the approximations are expected to hold.

\bibliography{biblio}

\end{document}